\documentstyle[aps,preprint]{revtex}
\begin{document}

%
%
%
%
%
\catcode`\@=11\relax
\newwrite\@unused
\def\typeout#1{{\let\protect\string\immediate\write\@unused{#1}}}
\typeout{psfig: version 1.1}
\def\psglobal#1{
\typeout{psfig: including #1 globally}
\immediate\special{ps:plotfile #1 global}}
\def\psfiginit{\typeout{psfiginit}
\immediate\psglobal{/usr/lib/ps/figtex.pro}}
%
%
\def\@nnil{\@nil}
\def\@empty{}
\def\@psdonoop#1\@@#2#3{}
\def\@psdo#1:=#2\do#3{\edef\@psdotmp{#2}\ifx\@psdotmp\@empty \else
    \expandafter\@psdoloop#2,\@nil,\@nil\@@#1{#3}\fi}
\def\@psdoloop#1,#2,#3\@@#4#5{\def#4{#1}\ifx #4\@nnil \else
       #5\def#4{#2}\ifx #4\@nnil \else#5\@ipsdoloop #3\@@#4{#5}\fi\fi}
\def\@ipsdoloop#1,#2\@@#3#4{\def#3{#1}\ifx #3\@nnil 
       \let\@nextwhile=\@psdonoop \else
      #4\relax\let\@nextwhile=\@ipsdoloop\fi\@nextwhile#2\@@#3{#4}}
\def\@tpsdo#1:=#2\do#3{\xdef\@psdotmp{#2}\ifx\@psdotmp\@empty \else
    \@tpsdoloop#2\@nil\@nil\@@#1{#3}\fi}
\def\@tpsdoloop#1#2\@@#3#4{\def#3{#1}\ifx #3\@nnil 
       \let\@nextwhile=\@psdonoop \else
      #4\relax\let\@nextwhile=\@tpsdoloop\fi\@nextwhile#2\@@#3{#4}}
\def\psdraft{
	\def\@psdraft{0}
}
\def\psfull{
	\def\@psdraft{100}
}
\psfull
\newif\if@prologfile
\newif\if@postlogfile
\newif\if@bbllx
\newif\if@bblly
\newif\if@bburx
\newif\if@bbury
\newif\if@height
\newif\if@width
\newif\if@rheight
\newif\if@rwidth
\newif\if@clip
\def\@p@@sclip#1{\@cliptrue}
\def\@p@@sfile#1{
		   \def\@p@sfile{#1}
}
\def\@p@@sfigure#1{\def\@p@sfile{#1}}
\def\@p@@sbbllx#1{
		\@bbllxtrue
		\dimen100=#1
		\edef\@p@sbbllx{\number\dimen100}
}
\def\@p@@sbblly#1{
		\@bbllytrue
		\dimen100=#1
		\edef\@p@sbblly{\number\dimen100}
}
\def\@p@@sbburx#1{
		\@bburxtrue
		\dimen100=#1
		\edef\@p@sbburx{\number\dimen100}
}
\def\@p@@sbbury#1{
		\@bburytrue
		\dimen100=#1
		\edef\@p@sbbury{\number\dimen100}
}
\def\@p@@sheight#1{
		\@heighttrue
		\dimen100=#1
   		\edef\@p@sheight{\number\dimen100}
}
\def\@p@@swidth#1{
		\@widthtrue
		\dimen100=#1
		\edef\@p@swidth{\number\dimen100}
}
\def\@p@@srheight#1{
		\@rheighttrue
		\dimen100=#1
		\edef\@p@srheight{\number\dimen100}
}
\def\@p@@srwidth#1{
		\@rwidthtrue
		\dimen100=#1
		\edef\@p@srwidth{\number\dimen100}
}
\def\@p@@sprolog#1{\@prologfiletrue\def\@prologfileval{#1}}
\def\@p@@spostlog#1{\@postlogfiletrue\def\@postlogfileval{#1}}
\def\@cs@name#1{\csname #1\endcsname}
\def\@setparms#1=#2,{\@cs@name{@p@@s#1}{#2}}
%
%
\def\ps@init@parms{
		\@bbllxfalse \@bbllyfalse
		\@bburxfalse \@bburyfalse
		\@heightfalse \@widthfalse
		\@rheightfalse \@rwidthfalse
		\def\@p@sbbllx{}\def\@p@sbblly{}
		\def\@p@sbburx{}\def\@p@sbbury{}
		\def\@p@sheight{}\def\@p@swidth{}
		\def\@p@srheight{}\def\@p@srwidth{}
		\def\@p@sfile{}
		\def\@p@scost{10}
		\def\@sc{}
		\@prologfilefalse
		\@postlogfilefalse
		\@clipfalse
}
%
%
\def\parse@ps@parms#1{
	 	\@psdo\@psfiga:=#1\do
		   {\expandafter\@setparms\@psfiga,}}
%
%
\newif\ifno@bb
\newif\ifnot@eof
\newread\ps@stream
\def\bb@missing{
	\typeout{psfig: searching \@p@sfile \space  for bounding box}
	\openin\ps@stream=\@p@sfile
	\no@bbtrue
	\not@eoftrue
	\catcode`\%=12
	\loop
		\read\ps@stream to \line@in
		\global\toks200=\expandafter{\line@in}
		\ifeof\ps@stream \not@eoffalse \fi
		\@bbtest{\toks200}
		\if@bbmatch\not@eoffalse\expandafter\bb@cull\the\toks200\fi
	\ifnot@eof \repeat
	\catcode`\%=14
}	
\catcode`\%=12
\newif\if@bbmatch
\def\@bbtest#1{\expandafter\@a@\the#1
\long\def\@a@#1
\long\def\bb@cull#1 #2 #3 #4 #5 {
	\dimen100=#2 bp\edef\@p@sbbllx{\number\dimen100}
	\dimen100=#3 bp\edef\@p@sbblly{\number\dimen100}
	\dimen100=#4 bp\edef\@p@sbburx{\number\dimen100}
	\dimen100=#5 bp\edef\@p@sbbury{\number\dimen100}
	\no@bbfalse
}
\catcode`\%=14
\def\compute@bb{
		\no@bbfalse
		\if@bbllx \else \no@bbtrue \fi
		\if@bblly \else \no@bbtrue \fi
		\if@bburx \else \no@bbtrue \fi
		\if@bbury \else \no@bbtrue \fi
		\ifno@bb \bb@missing \fi
		\ifno@bb \typeout{FATAL ERROR: no bb supplied or found}
			\no-bb-error
		\fi
		\count203=\@p@sbburx
		\count204=\@p@sbbury
		\advance\count203 by -\@p@sbbllx
		\advance\count204 by -\@p@sbblly
		\edef\@bbw{\number\count203}
		\edef\@bbh{\number\count204}
}
%
%
\def\in@hundreds#1#2#3{\count240=#2 \count241=#3
		     \count100=\count240	
		     \divide\count100 by \count241
		     \count101=\count100
		     \multiply\count101 by \count241
		     \advance\count240 by -\count101
		     \multiply\count240 by 10
		     \count101=\count240	
		     \divide\count101 by \count241
		     \count102=\count101
		     \multiply\count102 by \count241
		     \advance\count240 by -\count102
		     \multiply\count240 by 10
		     \count102=\count240	
		     \divide\count102 by \count241
		     \count200=#1\count205=0
		     \count201=\count200
			\multiply\count201 by \count100
		 	\advance\count205 by \count201
		     \count201=\count200
			\divide\count201 by 10
			\multiply\count201 by \count101
			\advance\count205 by \count201
		     \count201=\count200
			\divide\count201 by 100
			\multiply\count201 by \count102
			\advance\count205 by \count201
		     \edef\@result{\number\count205}
}
\def\compute@wfromh{
		\in@hundreds{\@p@sheight}{\@bbw}{\@bbh}
		\edef\@p@swidth{\@result}
}
\def\compute@hfromw{
		\in@hundreds{\@p@swidth}{\@bbh}{\@bbw}
		\edef\@p@sheight{\@result}
}
\def\compute@handw{
		\if@height 
			\if@width
			\else
				\compute@wfromh
			\fi
		\else 
			\if@width
				\compute@hfromw
			\else
				\edef\@p@sheight{\@bbh}
				\edef\@p@swidth{\@bbw}
			\fi
		\fi
}
\def\compute@resv{
		\if@rheight \else \edef\@p@srheight{\@p@sheight} \fi
		\if@rwidth \else \edef\@p@srwidth{\@p@swidth} \fi
}
%
\def\compute@sizes{
	\compute@bb
	\compute@handw
	\compute@resv
}
%
%
\def\psfig#1{\vbox {
	%
	\ps@init@parms
	\parse@ps@parms{#1}
	\compute@sizes
	\ifnum\@p@scost<\@psdraft{
		\typeout{psfig: including \@p@sfile \space }
		\special{ps::[begin] 	\@p@swidth \space \@p@sheight \space
				\@p@sbbllx \space \@p@sbblly \space
				\@p@sbburx \space \@p@sbbury \space
				startTexFig \space }
		\if@clip{
			\typeout{(clip)}
			\special{ps:: \@p@sbbllx \space \@p@sbblly \space
				\@p@sbburx \space \@p@sbbury \space
				doclip \space }
		}\fi
		\if@prologfile
		    \special{ps: plotfile \@prologfileval \space } \fi
		\special{ps: plotfile \@p@sfile \space }
		\if@postlogfile
		    \special{ps: plotfile \@postlogfileval \space } \fi
		\special{ps::[end] endTexFig \space }
		\vbox to \@p@srheight true sp{
			\hbox to \@p@srwidth true sp{
				\hfil
			}
		\vfil
		}
	}\else{
		\vbox to \@p@srheight true sp{
		\vss
			\hbox to \@p@srwidth true sp{
				\hss
				\@p@sfile
				\hss
			}
		\vss
		}
	}\fi
}}
\catcode`\@=12\relax

\title{  A New Approach of Fermion Field on Lattice }
\author{   Bo Feng$^a$, \footnote{Email address: lijm@itp.ac.cn}Jianming 
Li$^{a,b}$, 
Xingchang Song$^{a,b}$ \\}
       \address{ $a$ Department of Physics, Peking University, 
          Beijing 100871, China \\
           $b$ Institute of Theoretical Physics, Academia Sinica, Beijing 
100080, China} 
                   
\maketitle          
\begin{abstract}

 A new approach to formulate the fermion field on lattice is introduced by 
proposing
a new Dirac operator on lattice.
  This approach can eliminate the Fermion doubling problem, 
preserve the chiral symmetry and get
the same dispersion relation for both Fermion and Boson fields.Then  
the  Weinberg-Salam model on lattice may be formulated in this approach.
\end{abstract}


\newpage

    In lattice field theory, it   
  is well-known that the naive discretization of Fermion field suffers from 
the problem of Fermion doubling, while the scalar field doesn't.  The studies of 
Fermion species doubling problem persists a long duration since Wilson \cite{W1}
\cite{W2}.  
  To cure from this
Wilson \cite{W1}\cite{W2} added a new term, so-called the ``Wilson term'', 
to the lattice Fermion Lagrangian. This term kills the superfluous components 
of the Fermion field but 
breaks the chiral symmetry and leads to different dispersion relations for 
Fermions and Bosons. Kogut and Susskind\cite{S1}\cite{S2} proposed the 
``stagger model'', but it needs four generations of Fermions at least, 
and deals with the Fermions and Bosons on different footing.
Besides these two popular prescriptions, Drell et al\cite{D} developed another 
approach which preserves the chiral symmetry and at the same time correctly 
counts the number of Fermion states.  Their crucial
point was to introduce a lattice gradient operator which couples all lattice 
sites along a given direction instead of coupling only nearest-neighbor sites.
However, no convincing approach which can fit all the physical 
phenomena exists so far\cite{P}, 
such as we don't know how to reformulate the Weinberg-Salam model
on lattice. 
 
    In this paper, to overcome those difficulties  we propose to adopt
     a new approach in which a new Dirac 
operator is introduced. The square of this Dirac operator is the D'Alembertian 
operator, which recovers the relation in continue field theory.
By using this  approach, we can eliminate
the Fermion doubling problem and at the same time preserve the chiral symmetry.
Besides these, we can get the same dispersion relation for both Fermions and
Bosons.
\par
    First let us take a look upon how the Fermion-doubling problem arises. For
free scalar field $\phi(x)$,
the lattice action in the Euclidean space is (for more detail,
see \cite{Cheng} )
\begin{eqnarray}
\label{2}
S(\phi) = \sum_n \{ \frac{a^2}{2} \sum_{\mu}(\phi_{n+\mu}-\phi_n)^2
  +\frac{a^4}{2} m^2 \phi_n^2 \}. 
  \end{eqnarray}
 With the help of  displacement operator $R_{\mu}$, which is defined as
\begin{eqnarray}
\label{3}
R_{\mu} \phi_n= \phi_{n+\mu},
\end{eqnarray}
the free scalar action (\ref {2}) may be written as

\begin{eqnarray}
S(\phi)&= \sum_n \{ {a^2 \over 2} \sum_{\mu}[(\partial_\mu \phi_n]^2
  +\frac{a^4} {2} M^2 \phi_n^2 \} \nonumber \\
  &= \sum_n \{ {a^2\over 2} \phi^{\dag}\Box\phi_n
  +{a^4\over 2} M^2 \phi_n^2 \},
  \end{eqnarray}
  where the discrete derivative  $\partial_\mu={R_\mu-1 \over a}$  and 
  D'Alembertian operator  $\Box=\sum_{\mu}\frac {R_{\mu}+R_{-\mu}-2} {a^2}$.
Using the formula  
\begin{eqnarray}
\phi_n =[\int^{\pi\over a}_{\pi\over a}{dk^4\over (2\pi)^4}
 e^{ikna} \phi(k)] ,
 \end{eqnarray}
we can transfer (\ref{2}) into the momentum space form 
\begin{equation}
\label{5}
S(\phi)=\frac{1}{2} \int \frac{d^4 k}{ (2\pi)^4} \phi(-k) [ \sum_{\mu}
  \frac{4}{a^2} \sin^2(\frac{ak_{\mu}}{2})+M^2] \phi(k)             .
\end{equation}
The dispersion relation read from this equation is
\begin{equation}
\label{6}
S_{\phi}(k)= \sum_{\mu}\frac{4}{a^2} \sin^2(\frac{ak_{\mu}}{2})+M^2       .
\end{equation}

\par
The naive dicretization of free fermion field is usually taken as
    \begin{eqnarray}
\label{7}
S(\psi)&=&\sum_n \{ \frac{a^3}{2} \sum_{\mu} \bar{\psi}_n 
   (\stackrel{\rightarrow}{\not{\hskip -1.5mm {\cal D}}}
   -\stackrel{\leftarrow}{\not{\hskip -1.5mm {\cal D}}})\psi_n+Ma^4 \bar{\psi}_n 
\psi_n \}\nonumber\\
   &=&\sum_n \{ \frac{a^3}{2} \sum_{\mu} \bar{\psi}_n \gamma_{\mu}
   (R_{\mu}-R_{-\mu})\psi_n+Ma^4 \bar{\psi}_n \psi_n \} 
\end{eqnarray}
where ${\not{\hskip -1.5mm {\cal D}}}=\gamma_\mu \partial_\mu$ is the Dirac 
operator.
Transferring it into momentum space we get
\begin{eqnarray}
\label{8}
S(\psi)=\int \frac{d^4 k}{ (2\pi)^4} \bar{\psi}(-k) [i\sum_{\mu}
  \gamma_{\mu} \frac{\sin (ak_{\mu})}{a} +M] \psi(k)     .
\end{eqnarray}
From this equation we see the dispersion relation 
\begin{eqnarray}
\label{9}
S_{\psi}(k)= \sum_{\mu}\frac{1}{a^2} \sin^2(ak_{\mu}) + M^2
\end{eqnarray}
is different from (\ref{6}), and the doubling of  Fermionic degrees of freedom 
appears.

In above formulas, two key operators on lattice were introduced. The first is 
the 
Dirac operator $\not{\hskip -1.5mm {\cal D}}$, the second is the D'Alembertian 
operator $\Box$, which  
domain
 the dynamics of fields.
Now we may raise a question, why the lattice description of scalar field is 
successful, while the fermion field fails? One possible way to answer this 
question is to regard the lattice definition of D'Alembertian operator is 
``good'' but that of Dirac operator is ``bad''. So we advocated a new point that 
all the disaster of lattice Fermions come from the ``bad'' definition 
of Dirac operator. Therefore, to overcome the Fermionic difficulties
 we propose a new lattice 
Dirac operator,  which may shed the new light on studying
 lattice Fermion fields. 
  
Consulting the definition of the translation operator in the momentum 
space\footnote{The similar operator was also introduced in some papers at 
different angle\cite{Z,V}}.
\begin{eqnarray}
\label{12}
R_{\mu}\phi_n &=& R_{\mu} [\int^{\pi\over a}_{-\pi\over a}{dk^4\over (2\pi)^4}
 e^{ikna} \phi(k)] \nonumber \\
  &=& \phi_{n+\mu}=\int^{\pi\over a}_{-\pi\over a}{dk^4\over (2\pi)^4}
  e^{ikna} \psi(k) e^{ik_{\mu}a}
\end{eqnarray}
we can define an operator  $R_{\frac{\mu}{2}}$, corresponding to the 
``half-spacing translation'', as
\begin{eqnarray}
\label{13}
R_{\frac{\mu}{2}} \psi_n &:=& \int^{\pi\over a}_{-\pi\over a}
{dk^4\over (2\pi)^4} e^{ikna} \psi(k) e^{\frac{ik_{\mu}a}{2}};  
\nonumber  \\
R_{-\frac{\mu}{2}} \psi_n &:=& \int^{\pi\over a}_{-\pi\over a} 
{dk^4\over (2\pi)^4}e^{ikna}\psi(k) e^{-\frac{ik_{\mu}a}{2}}.
\end{eqnarray}

\par
    From the definition (\ref{13}) we can also see that $R_{\frac{\mu}{2}}$ 
and $R_{-\frac{\mu}{2}}$ are not ``local operators". When acting on some 
functions of lattice sites, they not only concern with the nearest neighbor 
sites but also all the sites in direction $\mu$ on the lattice. 
However, it 
is evident that the 
combination $\frac{ R_{\frac \mu 2}-R_{-\frac \mu 2} }{a}$ will 
approximate to $\partial_{\mu}$ when the lattice spacing $a \rightarrow 0$. So 
we can call the combination $\frac{R_{\frac{\mu}{2}}-R_{-\frac{\mu}{2}}}{a}$ 
the `` quasi-local operator". The quasi-local feature is the price for getting 
the above mentioned nice characters.

In this approach, the key point is to introduce the auxiliary operator
 $R_{\mu \over 2}$. Actually,
physical fields $\phi$ and $\psi$  distribute on the integer sites and the 
elementary
translation operator is $R_\mu$. To discuss the physical meaning of what we 
really
did, it is desirable to represent the auxiliary operator $R_{\mu\over2}$ in
terms of physical operator $R_\mu$. 

By the help of definition (\ref {13}), it is easy to show 
\begin{eqnarray}
R_{\mu\over 2}\psi_n&=&{a\over 2\pi} \sum_{m=-\infty}^{\infty}
\int_{-\pi\over a}^{\pi\over a} dk e^{ik\cdot (n-m+\frac 1 2 )a}\psi_m
\nonumber\\
&=&{1\over \pi}\sum_{m=-\infty}^{\infty} {(-1)^{n-m}\over n-m+\frac 1 2}
R^{m-n}_\mu\psi_n. 
\end{eqnarray}
So in terms of $R_\mu$,  the auxiliary operator $R_{\frac \mu 2}$ may be written 
as
\begin{eqnarray}\label{re1}
R_{\mu\over 2}={1\over \pi}\sum_{m=-\infty}^{\infty} {(-1)^{m}\over m+
\frac 1 2}
R^{m}_{-\mu}. 
\end{eqnarray}
Similarly, 
\begin{eqnarray}\label{re2}
R_{-{\mu\over 2}}=-{1\over \pi}\sum_{m=-\infty}^{\infty} {(-1)^{m}\over m-
\frac 1 2}
R^{m}_{-\mu}. 
\end{eqnarray}
From the above formulas, we see that the operators$R_{\frac \mu 2}$ and 
$R_{-\frac \mu 2}$ are well defined globally,  although  they were introduced in 
the first 
 Brillouin zone.

Using the expressions of operators $R_{\frac \mu 2}$, $R_{-\frac \mu 2 }$
in equations (\ref {re1}) and (\ref {re2})  we may prove the following relations 
hold,
\begin{eqnarray}
\label{b1}
R_{\frac{\mu}{2}}^2 &=& R_{\mu},  \\
R_{\frac{\mu}{2}} R_{-\frac{\mu}{2}} &=& R_{-\frac{\mu}{2}}R_{\frac{\mu}{2}}
              =I.
\end{eqnarray}

If  the Dirac operator is redefined  as 
\begin{eqnarray}
\label {dirac}
 {\not{\hskip -1.5mm {\cal D}}}=\gamma_\mu({R_{\mu\over 2}-R_{
\mu\over 2}\over a}),
\end{eqnarray}
by the help of equation (\ref{re1}) and (\ref {re2}), it may be written as
\begin{eqnarray}\label{re3}
{\not{\hskip -1.5mm {\cal D}}}={2\over \pi a}\sum_{m=1}^{\infty} {(-1)^{m} 
m\over 
m^2-
\frac 1 4}
\displaystyle\sum_\mu \gamma_\mu (R^{m}_{-\mu}-R_\mu^m). 
\end{eqnarray}
It is interesting to find that in this case we may prove the following equation 
exits 
\begin{eqnarray}\label{sq}
\Box={\not{\hskip -1.5mm {\cal D}}}^2,\end{eqnarray}
which recover the same relation of continuum field theory. In continuous limit
$a\rightarrow 0$, $\Box\rightarrow \sum_{\mu=1}^{4}\partial_\mu^2$, because the 
square root of D'Alembertian operator is unique in continuum theory,so by the 
help of equation (\ref {sq}), we show that the continuous limit of the new Dirac 
operator is just the ordinary Dirac operator.

Substitute the modified Dirac operator into  (\ref{7}), we may get the free 
fermion action  as 
\begin{eqnarray}
\label {mdirac}
S(\psi)=\sum_n \{ a^3 \frac 2 \pi \sum_{m=1}^{\infty}{(-1)^{m} m\over m^2-
\frac 1 4}\sum_\mu 
\bar{\psi}_n \gamma_{\mu}
   (R_{-\mu}^m-R_{\mu}^m)\psi_n+Ma^4 \bar{\psi}_n \psi_n \}.
\end{eqnarray}
which shows how the  fermion field on  non-neighbour sites correlated
each other. When $m$ is taken to $1$, (\ref {mdirac}) 
reduce to the form of   naive discretization
action (\ref {7}). 
Transform (\ref {mdirac}) into its   momentum space form, we have   
\begin{eqnarray}
\label{16}
S(\psi)&=&\int \frac{d^4 k}{ (2\pi)^4} \bar{\psi}(-k) [\sum_{\mu}
  \gamma_{\mu} (-\frac {4i} \pi )\sum_{m=1}^{\infty}{(-1)^{m} m\sin(mak_\mu) 
\over (m^2-
\frac 1 4  )a}+M] \psi(k)\\
&=&\int \frac{d^4 k}{ (2\pi)^4} \bar{\psi}(-k) [2i\sum_{\mu}
  \gamma_{\mu} {\Delta(ak_\mu)\over a} +M] \psi(k),
\end{eqnarray}
where function $\Delta(x)$ is defined as  
\begin{eqnarray}
\Delta(x)=\left\{ \begin{array}{cll}&\sin(\frac{x}{2}-p\pi) , 
~~~~~~~~~~&(2p-1)\pi<x<(2p+1)\pi\\
        &0,&x=(2p+1) \pi\end{array}\right..
\end{eqnarray} 
and may be illustrated by  following figure,\\
\psfig{figure=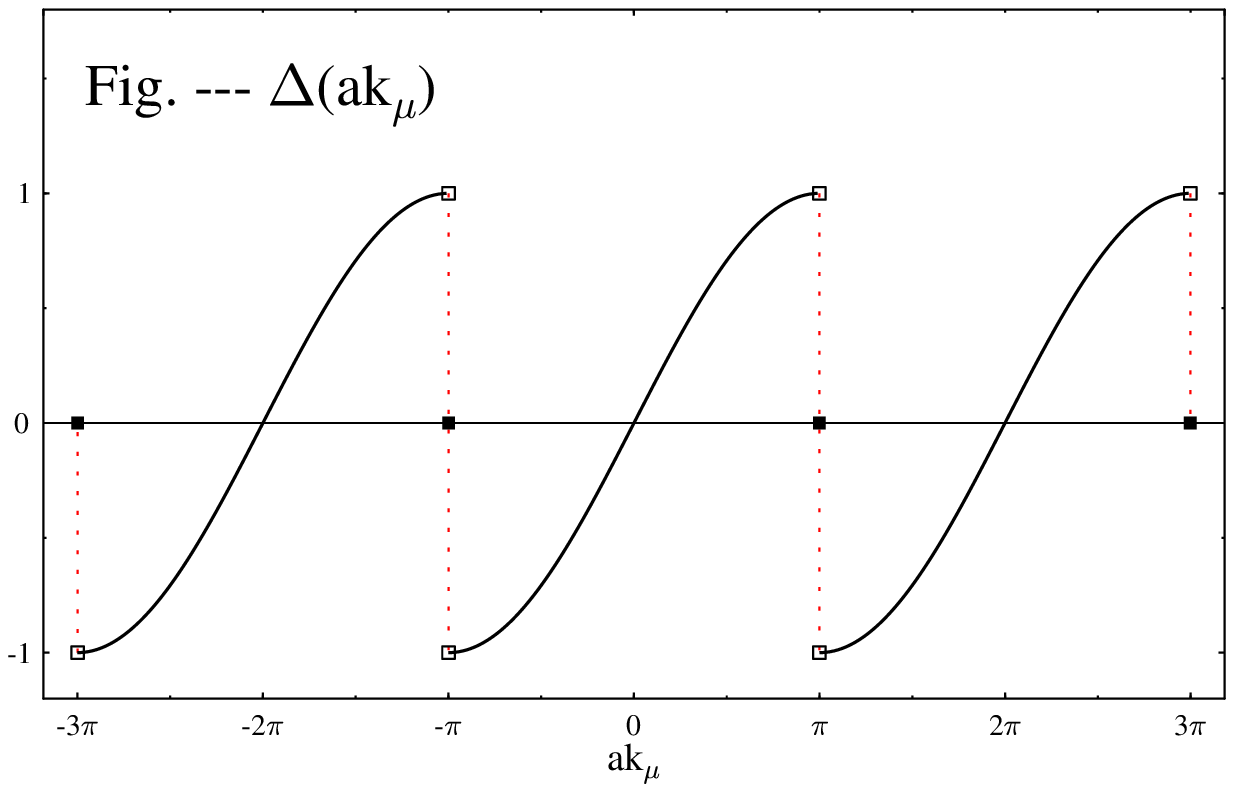,height=13cm,width=17cm}\\

 From this we see immediately that there is no double-counting anymore and
the Fermion field has the same dispersion relation as the one for 
scalar field (\ref{6}) one on the boundaries of the Brillouin zone $\Delta( 
ak_\mu)=0$. 
When $M=0$ this new formula (\ref{mdirac}) will preserve the chiral symmetry.

\par
    The next  work is to construct  the  gauge invariant action in this frame. 
    Under gauge transformation of lattice theory, $R_\mu$ will be changed 
    into its covariant form as
    \begin{eqnarray}
    &R_\mu \psi_n\longrightarrow U_{n,\mu} R_\mu \psi_n\nonumber\\
    &R_{-\mu} \psi_n\longrightarrow U^{\dag}_{n-\mu,\mu} R_\mu \psi_n
    \end{eqnarray}
    where $U_{n,\mu}$ are the known link variables.
 Then we have 
 \begin{eqnarray}
    &R_\mu^m \psi_n\longrightarrow (U_{n,\mu} R_\mu)^m \psi_n\nonumber\\
    &R_{-\mu}^m \psi_n\longrightarrow (U^{\dag}_{n-\mu,\mu} R_{-\mu})^m \psi_n
    \end{eqnarray}
    Expanding the exponential, one gets
    \begin{eqnarray}
    &(U_{n,\mu} R_\mu)^m=U_{n,\mu}U_{n+\mu,\mu}\cdots U_{n+(m-1)\mu,\mu} R_\mu^m
   \nonumber\\
   &(U^{\dag}_{n-\mu,\mu} 
R_{-\mu})^m=U^{\dag}_{n-\mu,\mu}U^{\dag}_{n-2\mu,\mu}\cdots 
U_{n-m\mu,\mu} R_{-\mu}^m
    \end{eqnarray}
 
 Now we can build a gauge invariant toy model including a Fermion field $\psi$, 
scalar field $\phi$ and gauge field $U_{n,\mu}$, where $U_{n,\mu}$  is the link 
variable of gauge 
group $G$. The action of the toy model may be written out as following
 \begin{eqnarray}
 S=S_{Fermion}+S_{Scalar}+ S_{gauge}
 \end{eqnarray}
 where 
 \begin{eqnarray}
 S_{Fermion}&=&\sum_n \{ a^3 \frac 2 \pi \sum_{m=1}^{\infty}{(-1)^{m} m\over 
m^2-
\frac 1 4}\sum_\mu 
\bar{\psi}_n \gamma_{\mu}
  [(U^{\dag}_{n,\mu} R_{-\mu})^m-(U_{n,\mu} R_{\mu})^m)]\psi_n\nonumber\\
  &&+\lambda 
   (\bar{\psi}^L_n\phi_n \psi^R_n+\bar{\psi}^R_n\phi^{\dag}_n \psi^L_n 
\},\nonumber
\end{eqnarray}
\begin{eqnarray}
S_{Scalar}=\sum_n \{ \frac{a^2}{2} \phi^{\dag}(U_{n,\mu} R_\mu 
+U^{\dag}_{n,\mu} R_{-\mu}-2]\phi_n]^2
  +\frac{a^4}{2} m^2 \phi_n^2 \},
  \end{eqnarray}
  \begin{eqnarray}
  S_{Gauge}=\frac 1 {g^2} \sum_{Plaqu}[1-\frac 1 2 (U_P+U_P^{\dag})]\nonumber,
  \end{eqnarray}
  where $\psi^L$ and $\psi^R$ are the left and right hand Fermion respectively.

\par
  With all the previous points in mind, a construction for the $SU(2)\times 
U(1)$
  electroweak theory on lattice is now available. Consider just the leptonic 
  first generation section: electron and electron neutrino
  and define the fields as follows:
  \begin{eqnarray}
  &L=\left(\begin{array}{cl}\nu_L\\e_L \end{array}\right),  &R=e_R\nonumber\\
  &U^L_{n,\mu}=U_{n,\mu}V_{\mu},  &U^R_{n,\mu}=V_{n,\mu}^{\prime}
  \end{eqnarray}
  where  $U_{n,\mu}\exp^{iga\tau\cdot A_{n,\mu}}$ is $SU(2)$ gauge field and 
  $V_{n,\mu}=\exp^{-\frac 1 2 ig^{\prime}a B_{n,\mu}}$,$V_{n,\mu}^{\prime}=
  \exp^{-ig^{\prime}aB_{n,\mu}}$ are $U(1)$ gauge field.
  Because the action of scalar field and gauge field parts are easy to write 
out, we
  only give  the action of fermionic part here,
  \begin{eqnarray}
 S_{Fermion}&=&\sum_n \{ a^3 \frac 2 \pi \sum_{m=1}^{\infty}{(-1)^{m} m\over 
m^2-
\frac 1 4}\sum_\mu 
\bar{L}_n \gamma_{\mu}
   [(U^L_{n,\mu}R_{-\mu})^m-(U^L_{n,\mu}R_{\mu})^m)L_n\nonumber\\
   &&+\sum_n \{ a^3 \frac 2 \pi \sum_{m=1}^{\infty}{(-1)^{m} m\over 
m^2-
\frac 1 4}\sum_\mu 
\bar{R}_n \gamma_{\mu}
   [(G^R_{n,\mu}R_{-\mu})^m-(U^R_{n,\mu}R_{\mu})^m)R_n\nonumber\\
   &&+\lambda a^4 
   (\bar{L}_n \phi_n R_n +\bar{R}_n\phi^{\dag}_n L_n)\},
\end{eqnarray}
where $\lambda$ is Yukawa coupling constant and $\phi=\left(\begin{array}{cl}
\phi_+ \\\phi_0\end{array}\right)$ is the Higgs doublet.

\par
    In summary, in this paper we have proposed to introduce the  Dirac operator
    ${\not{\hskip -1.5mm {\cal D}}}={2\over \pi a 
}\displaystyle\sum_{m=1}^{\infty} 
{(-1)^{m} m\over m^2-
\frac 1 4}
(R^{m}_{-\mu}-R_\mu^m)$  in
lattice theory, which shares the property ${\not{\hskip -1.5mm {\cal 
D}}}^2=\Box$. By doing so, we can eliminate the Fermion-doubling problem,
get the same dispersion relation for both Boson and Fermion fields
and preserve the chiral symmetry. We can also construct the gauge theory
in the new frame. Because of these good characters, we can discuss chiral
model  which is difficult to touch upon in normal lattice theory and we 
completed the construction of Weinber-Salam Electro-Weak model on lattice.
 Modifying the Dirac operator is not new in lattice theory, actually, the 
 introduction of Wilson fermion may also be regarded as another kind of 
modifying method 
 in some sense.

\par
\bigskip
The authors would like to thank Dr. Zenkin and Dr.A.Polychronakos for their useful 
discussions and   
Jianming Li also want to give thanks to Professors H.Y Guo, Ke Wu and C.P. Sun 
for 
their stimulate discussions and help. 

This work is supported in part by the National PanDeng (Clime Up) Plan
and the Chinese National Science Foundation and in part by the Chinese 
Postdoctoral Foundation.

\end{document}